\documentclass[aps,preprint,showkeys, showpacs]{revtex4}

\def\be{\begin{equation}}
\def\ee{\end{equation}}
\def\bea{\begin{eqnarray}}
\def\eea{\end{eqnarray}}

\begin{document}
\preprint{SOGANG-HEP 285/01}
\date{\today}

\title{Global Embeddings of Two-dimensional Dilatonic Black Holes}

\author{Soon-Tae \surname{Hong},\footnote{electronic address: sthong@sogang.ac.kr} 
Won Tae \surname{Kim},\footnote{electronic address: wtkim@ccs.sogang.ac.kr}
John J. \surname{Oh},\footnote{electronic address: john5@string.sogang.ac.kr} and 
Young-Jai \surname{Park}\footnote{electronic address: yjpark@sogang.ac.kr}}
 
\affiliation{Department of Physics and Basic Science Research Institute,
Sogang University, C.P.O. Box 1142, Seoul 100-611, Korea} 

\begin{abstract}
We obtain minimal (2+1)- and (2+2)-dimensional global flat embeddings of 
uncharged and charged dilatonic black holes in (1+1)-dimensions. Moreover, we obtain the Hawking temperatures and the black 
hole temperatures of these dilatonic black holes.  However, even though the 
minimal flat embedding structures are mathematically meaningful, through 
these minimal embeddings, the proper entropies are shown to be unattainable, 
in contrast to the cases of other black holes in (2+1) or much higher 
dimensions.  
\end{abstract}
\keywords{superstring theory, global flat embedding}
%PACS number(s): 04.70.Dy, 04.62.+v, 04.20.Jb, 11.25.-w
\pacs{PACS number(s): 04.70.Dy, 04.62.+v, 04.20.Jb, 11.25.-w}
\maketitle

%%%%%%%%%%%%%%%%%%%%%%%
\section{Introduction \hfil{}}\label{int}
%%%%%%%%%%%%%%%%%%%%%%%

There has been tremendous progress in the study of two-dimensional black
holes\cite{bhs} and string theory\cite{witten91}.  It is also well known that 
in the string theory, a U-duality exists between two-dimensional
dilatonic black holes\cite{nappi92,nappi92mod,gibbons92,hyun98} and
five-dimensional ones. On the other hand, it is well-known that a thermal Hawking effect on a 
curved manifold\cite{hawk75} can be viewed at as an Unruh effect\cite{unr} 
in a higher-dimensional flat spacetime.  Following the global 
embedding Minkowski space (GEMS) approach\cite{kasner}, several 
authors\cite{deser97,des,kps99,kps00} recently have shown that this approach 
could yield a unified derivation of temperature for various curved manifolds,
such as the rotating Banados-Teitelboim-Zanelli 
(BTZ) manifold\cite{btz1,cal,kps,mann93,jkps}, Schwarzschild
manifold\cite{sch} together with its anti-de Sitter (AdS) extension,
the Reissner-Nordstr\"om (RN) manifold\cite{rn} and the RN-AdS
manifold\cite{kps99}.   

Historically, the higher-dimensional global flat 
embeddings of black-hole solutions have been subjects of great interest to 
mathematicians, as well as physicists.  In differential geometry, it is well-known that the four-dimensional Schwarzschild metric is not embedded 
in $R^{5}$\cite{spivak75}.  Recently, Deser and Levin first obtained 
(5+1)-dimensional global flat embeddings of the (3+1) Schwarzschild black-hole 
solution\cite{deser97}.  The (3+1)-dimensional RN-AdS, RN, and 
Schwarzschild-AdS black holes are also shown to be embedded in
(5+2)-dimensional GEMS manifolds\cite{kps99}. On the other hand, very
recently, the brane metric has also been embedded in six
dimensions.\cite{brane} 

Moreover, static, rotating, and charged (2+1)-dimensional BTZ AdS 
black holes are shown to have (2+2), (2+2), and (3+3) GEMS 
structures\cite{deser97,kps00}, respectively, while the static, rotating, and 
charged (2+1)-dimensional dS black holes are shown to have (3+1), (3+1) and 
(3+2) GEMS structures, respectively\cite{kps00}.  Very recently, we have 
obtained (3+1) and (3+2) GEMS of uncharged and charged (2+1) black 
strings, respectively\cite{hong01prd}.  Until now, we have analyzed the GEMS 
structure of the black hole and black strings in (2+1) and (3+1) dimensions.  
It is now interesting to study the geometry of (1+1)-dimensional
dilatonic black-hole solutions in the GEMS approach to directly yield
their minimal flat embeddings. 

In this paper, we will analyze the Hawking and Unruh effects of (1+1)-dimensional dilatonic black holes\cite{nappi92,nappi92mod,gibbons92} in the 
framework of the GEMS scheme.  In Section \ref{II}, we will briefly recapitulate 
two-dimensional dilatonic black holes\cite{nappi92,nappi92mod,gibbons92} 
associated with type IIA strings to yield asymptotically flat 
two-dimensional dilatonic black holes.  In Section \ref{III}, we will obtain the 
minimal GEMS structures of uncharged and charged two-dimensional dilatonic 
black holes and their corresponding Hawking temperatures.  In Section \ref{IV}, we 
will discuss the entropies of dilatonic black holes and the embedding 
constraints in the framework of the GEMS scheme. 

%%%%%%%%%%%%%%%%%%%%%%%%%%%%%%%%%%%%%%%%%%%%%%%%%%%%%%%%%%%%%%%%%
\section{Type IIA string theory and two-dimensional black holes \hfil{}}\label{II}
%%%%%%%%%%%%%%%%%%%%%%%%%%%%%%%%%%%%%%%%%%%%%%%%%%%%%%%%%%%%%%%%%

In this section, we briefly recapitulate two-dimensional dilatonic black 
holes\cite{nappi92,nappi92mod,gibbons92} associated with type IIA string 
theories and their compactification to five dimensions whose metric is the 
product of a three-sphere and an asymptotically flat two-dimensional 
geometry.  The ten-dimensional type IIA superstring solution consists of a 
solitonic NS 5-brane wrapping around the compact coordinates, say, 
$x_{5}$, $x_{i}$ $(i=6,7,8,9)$ and a fundamental string wrapping around 
$x_{5}$, and a gravitational wave propagating along $x_{5}$.  In the 
string frame, the 10-metric, dilaton and 2-form field $B$ are given 
as\cite{horowitz962,tsey96,mal96} 
\bea
ds^{2}&=&-(H_{1}K)^{-1}fdt^{2}+H_{1}^{-1}K(dx_{5}-(K^{\prime -1}-1)dt)^{2}
+H_{5}(f^{-1}dr^{2}+r^{2}d\Omega_{3}^{2})+dx_{i}dx^{i},
\nonumber\\
e^{-2\phi}&=&H_{1}H_{5}^{-1},~~
B_{05}=H_{1}^{\prime -1}-1+\tanh\alpha,~~
B_{056789}=H_{5}^{\prime -1}-1+\tanh\beta,
\eea
where $r^{2}=x_{1}^{2}+\cdots+x_{4}^{2}$ and 
\bea
H_{1}&=&1+\frac{r_{0}^{2}\sinh^{2}\alpha}{r^{2}},~~
H_{5}=1+\frac{r_{0}^{2}\sinh^{2}\beta}{r^{2}},~~
K=1+\frac{r_{0}^{2}\sinh^{2}\gamma}{r^{2}},
\nonumber\\
H_{1}^{\prime -1}&=&1-\frac{r_{0}^{2}\sinh\alpha\cosh\alpha}{r^{2}H_{1}},~~
K^{\prime -1}=1-\frac{r_{0}^{2}\sinh\gamma\cosh\gamma}{r^{2}K},~~
f=1-\frac{r_{0}^{2}}{r^{2}}.
\eea
Here, the $B_{05}$ component of the Neveu-Schwarz 2-form $B$ is the electric 
field of the fundamental strings, and $B_{056789}$ is the electric field dual to the 
magnetic field of the 5-brane with components $B_{ij}$.  Exploiting
the dimensional reduction in $x_{5}$, $x_{i}$ ($i=6,7,8,9$) directions in the 
Einstein frame, one can obtain the five-dimensional black hole 
metric\cite{horowitz962,tsey96}
\be
ds^{2}=-(H_{1}H_{5}K)^{-2/3}fdt^{2}+(H_{1}H_{5}K)^{1/3}
(f^{-1}dr^{2}+r^{2}d\Omega_{3}^{2}).
\ee

On the other hand, performing an $T_{5}ST_{6789}ST_{5}$ transformation with 
the $T$-duality $T_{ij\dots}$ along the $ij\dots$ directions and the 
$S$-duality $S$ of type IIB string\cite{berg95} and then performing an SL(2,R) 
coordinate transformation associated with the O(2,2) $T$-duality 
group, one can also obtain the 5-metric
\bea
ds^{2}&=&-(H_{1}\bar{H}_{5})^{-1}fdt^{2}+H_{1}^{-1}\bar{H}_{5}(dx_{5}
-(\bar{H}_{5}^{-1}-1)dt)^{2}
\nonumber\\
& &+K(f^{-1}dr^{2}+r^{2}d\Omega_{3}^{2})(f^{-1}dr^{2}+r^{2}d\Omega_{3}^{2})
\label{ds1022}
\eea
with 
\be
\bar{H}_{5}=\frac{r_{0}^{2}}{r^{2}}.
\ee
Performing the same set of $S$ and $T$ transformations reversely, one can 
obtain 
\bea
ds^{2}&=&-(H_{1}^{-3}\bar{H}_{5})^{-1/4}K^{-1}fdt^{2}+(H_{1}^{-3}\bar{H}_{5}
)^{-1/4}K(dx_{5}-K^{\prime -1}-1)dt)^{2}
\nonumber\\
& &+(H_{1}\bar{H}_{5}^{3})^{1/4}(f^{-1}dr^{2}+r^{2}d\Omega_{3}^{2})
+(H_{1}\bar{H}_{5}^{-1})^{1/4}dx_{i}dx^{i},
\label{ds103}\\
e^{-2\phi}&=&\frac{r^{2}}{r_{0}^{2}}+\sinh^{2}\alpha,
\label{dila5}
\eea
which, after exploiting dimensional reduction in the $x_{5}$, $x_{i}$ 
($i=6,7,8,9$) directions with $\alpha=\gamma$, yield the five-dimensional 
black hole metric\cite{teo98}
\bea
ds^{2}&=&-\left(1-\frac{r_{0}^{2}}{r^{2}}\right)\left(1+\frac{r_{0}^{2}\sinh^{2}\alpha}{r^{2}}
\right)^{-2}dt^{2}+\left(\frac{r^{2}}{r_{0}^{2}}-1\right)^{-1}dr^{2}
+r_{0}^{2}d\Omega_{3}^{2},
\label{ds533}
\eea
and the dilaton which is trivially invariant under the 
dimensional reduction to yield the same result as in Eq. (\ref{dila5}).  Here, 
one notes that the metric in Eq. (\ref{ds533}) is the product of two 
completely decoupled parts, namely, a three-sphere and an asymptotically 
flat a two-dimensional geometry which describes two-dimensional charged 
dilatonic black hole. Now, introducing a new variable $x$ and the
parameters $m$, $q$ and $Q$ such that 
\be
e^{Qx}=2\left(\frac{r^{2}}{r_{0}^{2}}+\sinh^{2}\alpha\right)
(m^{2}-q^{2})^{1/2},~~~Q=\frac{2}{r_{0}},
\label{newv}
\ee
where $m$ and $q$ are the mass and the charge of the dilatonic black hole, one 
can obtain the well-known two-dimensional charged dilatonic black 
hole metric\cite{nappi92,nappi92mod}
\be
ds^{2}=-(1-2me^{-Qx}+q^{2}e^{-2Qx})dt^{2}
+(1-2me^{-Qx}+q^{2}e^{-2Qx})^{-1}dx^{2},
\label{nappimetric}
\ee
which will be discussed in the framework of the GEMS scheme in the next 
section.

%%%%%%%%%%%%%%%%%%%%%%%%%%%%%%%%%%%%%%%%%%%%%%%%%%
\section{Minimal GEMS structures of two-dimensional dilatonic black
  holes \hfil{}}\label{III}
%%%%%%%%%%%%%%%%%%%%%%%%%%%%%%%%%%%%%%%%%%%%%%%%%%

Now, we consider the two-dimensional charged dilatonic black hole 
having the 2-metric\cite{nappi92,nappi92mod}
\be
ds^{2}=-N^{2}dt^{2}+N^{-2}dx^{2},
\label{2metric}
\ee
where the lapse function is given as
\be
N^{2}=1-2m e^{-Qx}+q^{2}e^{-2Q x},
\label{lapse}
\ee
from which one can obtain the horizons $x_{H}$ and $x_{-}$ in terms of the 
mass $m$ and the charge $q$,
\bea
e^{Q x_{H}}&=&m+(m^{2}-q^{2})^{1/2},
\nonumber\\
e^{Q x_{-}}&=&m-(m^{2}-q^{2})^{1/2}.
\label{hor}
\eea
By using these relations, one can rewrite the lapse function as 
\be
N^{2}=\left(1-e^{-Q(x-x_{H})}\right)
\left(1-e^{-Q(x-x_{-})}\right).
\ee

In order to construct the GEMS structures of the two-dimensional dilatonic 
black hole, we first consider the uncharged dilatonic black-hole 2-metric
\be
ds^{2}=-\left(1-2m e^{-Qx}\right)dt^{2}
+\left(1-2me^{-Qx}\right)^{-1}dx^{2}.
\label{dsdbh0}
\ee
Making an ansatz of two coordinates $(z^{0}, z^{1})$ in Eq. (\ref{zzz0})
to yield 
\bea
-(dz^0)^2+(dz^1)^2&=&-(1-e^{-Q(x-x_{H})})dt^{2}
+\frac{e^{-2Q(x-x_{H})}}{1-e^{-Q(x-x_{H})}}dx^{2}
\nonumber\\
&=&ds^{2}-(1+e^{-Q(x-x_{H})})dx^{2}\equiv ds^{2}-(dz^{2})^{2},
\label{dz0123d}
\eea
we obtain the (2+1)-dimensional minimal GEMS black hole metric
$ds^{2}=-(dz^{0})^2+(dz^{1})^2+(dz^{2})^2$ given by the 
coordinate transformations 
\bea
z^{0}&=&k_{H}^{-1}\left(1-e^{-Q (x-x_{H})}\right)^{1/2}
\sinh k_{H}t, \nonumber \\
z^{1}&=&k_{H}^{-1}\left(1-e^{-Q (x-x_{H})}\right)^{1/2}
\cosh k_{H}t, \nonumber \\
z^{2}&=&\frac{2}{Q}\left((1+e^{-Q (x-x_{H})})^{1/2}
+\frac{1}{2}\ln \frac{(1+e^{-Q (x-x_{H})})^{1/2}-1}
{(1+e^{-Q (x-x_{H})})^{1/2}+1}\right),
\label{zzz0}
\eea
where $k_{H}=\frac{Q}{2}$ is the surface gravity and $x_{H}$ is given by 
Eq. (\ref{hor}) with $q=0$.  For the trajectories, which follow the 
Killing vector $\xi=\partial_{t}$ on the uncharged two-dimensional dilatonic 
black-hole manifold described by $(t,x)$, one can obtain the constant 
2-acceleration,
\be
a_{2}=\frac{Qe^{-Q (x-x_{H})}}{2(1-e^{-2Q (x-x_{H})})^{1/2}},
\ee
from the definition of the acceleration in $n$-dimensional spacetimes,
$a_{n}=\sqrt{a_{\alpha}a^{\alpha}}$, where $a_{\alpha} = {\xi^{\mu}\nabla_{\alpha}\xi_{\mu}}/|\xi|^2$.
Moreover, we can obtain the Hawking temperature and the black-hole
temperature :
\bea
T_{H}&=&\frac{a_{3}}{2\pi}=\frac{Q}{4\pi}
\frac{1}{\left(1-e^{-Q (x-x_{H})}\right)^{1/2}},
\nonumber\\
T&=&(-g_{00})^{1/2}T_{H}=\frac{Q}{4\pi},
\label{t0}
\eea
where $a_{3}$ is a 3-acceleration in embedded higher-dimensional
spacetimes, which is given by $\kappa_{H}\sqrt{-g^{tt}}$.
Here, one notes that the above Hawking temperature is also given by the 
relation\cite{hawk75}
\be
T_{H}=\frac{1}{2\pi}\frac{k_{H}}{(-g_{00})^{1/2}}.
\ee

Next, we consider the charged dilatonic black hole whose 2-metric is given by 
Eqs. (\ref{2metric}) and (\ref{lapse}). Similarly to the uncharged case, 
after some algebra, we arrive at the (2+2) GEMS metric 
$ds^{2}=-(dz^{0})^2+(dz^{1})^2+(dz^{2})^2-(dz^{3})^{2}$ for the charged 
two-dimensional dilatonic black hole given by the coordinate transformations 
\bea
z^{0}&=&k_{H}^{-1}\left(1-e^{-Q (x-x_{H})}\right)^{1/2}
\left(1-e^{-Q (x-x_{-})}\right)^{1/2}
\sinh k_{H}t, \nonumber \\
z^{1}&=&k_{H}^{-1}\left(1-e^{-Q (x-x_{H})}\right)^{1/2}
\left(1-e^{-Q (x-x_{-})}\right)^{1/2}
\cosh k_{H}t, \nonumber \\
z^{2}&=&\frac{2}{Q}\left(e^{Q(x_{H}-x_{-})/2}\tan^{-1}
(e^{-Q (x_{H}-x_{-})}F)^{1/2}
+\frac{1}{2}\ln \frac{F^{1/2}-1}{F^{1/2}+1}\right)
\nonumber\\
z^{3}&=&\frac{2e^{-3Q(x-x_{H})/2}e^{-Q(x-x_{-})/2}}
{Q(e^{-Q(x-x_{H})}-e^{-Q(x-x_{-})})},
\label{dsch}
\eea
where the surface gravity $k_{H}$ and $F$ are given as 
\bea
k_{H}&=&\frac{Q}{2}(1-e^{-Q(x_{H}-x_{-})}),
\label{kh}\\ 
F&=&\frac{1+e^{-Q(x-x_{H})}}{1-e^{-Q(x-x_{-})}}.
\eea
Here, one can easily check that, in the uncharged limit ($q\rightarrow 0$ or $e^{Qx_{-}}\rightarrow 0$), the above coordinate 
transformations reduce exactly to the previous ones in Eq. (\ref{zzz0}) for the 
uncharged dilatonic black hole case.\footnote{For the case of $z^{2}$,
  one needs to exploit L'Hospital's rule to explicitly see the
  uncharged limit.}.  For the trajectories, which follow the Killing
vector $\xi=\partial_{t}$ on the charged dilatonic black-hole manifold
described by $(t,x)$, one can obtain the constant 2-acceleration,
\be
a_{2}=\frac{Qm e^{-Q (x-x_{H})}}
       {\left(1-e^{-Q (x-x_{H})}\right)^{1/2}
        \left(1-e^{-Q (x-x_{-})}\right)^{1/2}
        (m+(m^{2}-q^{2})^{1/2}) },
\ee
and the Hawking temperature and the black-hole temperature,
\bea
T_{H}&=&\frac{a_{4}}{2\pi}=\frac{Q}{4\pi}
\frac{1-e^{-Q(x_{H}-x_{-})}}
{\left(1-e^{-Q(x-x_{H})}\right)^{1/2}
\left(1-e^{-Q(x-x_{-})}\right)^{1/2}},
\nonumber\\
T&=&(-g_{00})^{1/2}T_{H}=\frac{Q}{4\pi}
(1-e^{-Q(x_{H}-x_{-})})^{1/2},
\eea
where $a_{4}$ is an acceleration in embedded four-dimensional spacetimes.
%%%%%%%%%%%%%%%%%%%%%%%%%%%%%%%%%%%%%%%%%%%%%%%%%%
\section{Entropies of two-dimensional dilatonic black holes \hfil{}}\label{IV}
%%%%%%%%%%%%%%%%%%%%%%%%%%%%%%%%%%%%%%%%%%%%%%%%%%

In this section, we consider the entropies of the dilatonic
black holes in the framework of the GEMS scheme.  For the uncharged
case, the Rindler horizon condition $(z^1)^2- (z^0)^2 = 0$ implies $r=r_H$, 
and the remaining embedding constraints yield $z^{1}=f(r)$ where 
$f(r)$ can be read off from Eq. (\ref{zzz0}).  The area of the Rindler 
horizon\cite{gib77} then seems to yield the entropy of the uncharged 
dilatonic black hole :
\be
S=\frac{1}{4G_{3}}\int{\rm d}z^{2}\delta(z^{2}-f(r))=\frac{1}{4G_{3}},
\label{entropy0}
\ee
where we have explicitly included the constant of proportionality 
$1/4G_{3}$\cite{nature}.  However, the entropy in Eq. (\ref{entropy0}) is not 
equivalent to the previous result in Refs.\cite{nappi92mod} and\cite{teo98} since we still have the Newton constant $G_{3}$ instead of $G_{2}$.  Moreover, in 
defining the entropy (\ref{entropy0}), we have used an improper constraint 
condition, $\delta(z^{2}-f(r))$, since one needs at least one nontrivial 
constraint describing a relation among the GEMS coordinates and the
constraint in Eq. (\ref{entropy0}) cannot yield any relation between the 
coordinates in Eq. (\ref{zzz0}).  In that sense, one cannot 
obtain the desired proper entropy in the minimal higher-dimensional 
embeddings of the dilatonic black hole.  In order to avoid these 
difficulties, as a plausible candidate, one could consider other
higher-dimensional embeddings such as (3+1)-dimensional GEMS : 
\bea
z^{0}&=&k_{H}^{-1}\left(1-e^{-Q (x-x_{H})}\right)^{1/2}
\sinh k_{H}t, \nonumber \\
z^{1}&=&k_{H}^{-1}\left(1-e^{-Q (x-x_{H})}\right)^{1/2}
\cosh k_{H}t, \nonumber \\
z^{2}&=&x,\nonumber\\
z^{3}&=&\frac{2}{Q}e^{-Q (x-x_{H})/2}.
\label{zzz04}
\eea
Then, using the above nonminimal GEMS and using the 
relation $G_{4}=V_{2}G_{2}=2G_{2}/Q$, where $V_{2}$ is a compact volume, 
$V_{2}=2/Q$, given along $z^{2}$ only, one could obtain the desired entropy :
\bea
S&=&\frac{1}{4G_{4}}\int{\rm d}z^{2}{\rm d}z^{3}\delta(z^{3}-\frac{2}{Q}
e^{-Q (z^{2}-x_{H})/2})
\nonumber\\
&=&\frac{1}{4G_{4}}\int_{0}^{\frac{2}{Q}}{\rm d}z^{3}
=\frac{1}{4G_{2}},
\label{entropy000}
\eea
which is consistent with the previous result in 
Refs.\cite{nappi92mod} and\cite{teo98}. A similar situation happens in
the case of the charged two-dimensional dilatonic black hole.

For a charged dilatonic black hole case, for instance, one could consider a 
(3+2) nonminimal GEMS metric
$ds^{2}=-(dz^{0})^2+(dz^{1})^2+(dz^{2})^2+(dz^{3})^{2}-(dz^{4})^{2}$
given by the coordinate transformations  
\bea
z^{0}&=&k_{H}^{-1}\left(1-e^{-Q (x-x_{H})}\right)^{1/2}
\left(1-e^{-Q (x-x_{-})}\right)^{1/2}\sinh k_{H}t, \nonumber \\
z^{1}&=&k_{H}^{-1}\left(1-e^{-Q (x-x_{H})}\right)^{1/2}
\left(1-e^{-Q (x-x_{-})}\right)^{1/2}\cosh k_{H}t, \nonumber \\
z^{2}&=&x,\nonumber\\
z_{3}&=&\frac{2}{Q} (1+e^{Q(x_{H}-x_{-})})^{1/2}\sin^{-1}e^{-Q(x-x_{-})/2},
\nonumber \\
z^{4}&=&\frac{2e^{-3Q(x-x_{H})/2}e^{-Q(x-x_{-})/2}}
{Q(e^{-Q(x-x_{H})}-e^{-Q(x-x_{-})})}.
\label{zzz05}
\eea
Here, one can also check that, in the uncharged limit, $q\rightarrow 0$, the 
above coordinate transformations reduce exactly to the previous ones
in Eq. (\ref{zzz04}) for the uncharged dilatonic black-hole case.  However, with this 
simplest nonminimal GEMS structure, one cannot obtain a proper entropy 
consistent 
with previous results\cite{nappi92mod, teo98}.  How many higher 
dimensions is required to fix the GEMS structure of this charged case 
to yield the proper entropy is highly nontrivial. 

%%%%%%%%%%%%%%%%%%%%%%%%%%%%%%%%%%%%%%%%%%%%%%%
\section{Conclusions \hfil{}}\label{con}
%%%%%%%%%%%%%%%%%%%%%%%%%%%%%%%%%%%%%%%%%%%%%%%

In conclusion, we have investigated the higher-dimensional global 
flat embeddings of (1+1) uncharged and charged dilatonic black holes.  These 
two-dimensional dilatonic black holes are shown to be minimally embedded in 
(2+1) and (3+1) dimensions for the uncharged and the charged two-dimensional dilatonic black holes, respectively.  In these 
minimal GEMS, we have obtained the 2-accelerations, the Hawking temperatures, 
and the black-hole temperatures, which are independent of the dimensionalities 
of the GEMS structures since they are calculated only in terms of the 
original metrics (or more practically in terms of the lapse functions), regardless of the GEMS coordinate transformations.  However, even though the minimal GEMS 
structures are mathematically meaningful, one has difficulties 
in deriving the GEMS coordinate transformations to yield the proper 
entropies since the entropies of the dilatonic black holes have nontrivial $G_{n}$ factors which 
are associated with the U-duality structure involved in type IIA string 
theory and depend on the dimensionalities of the GEMS structures.  In fact, we 
have succeeded in obtaining a (3+1) GEMS structure for the uncharged 
dilatonic black hole that yielded an entropy consistent with previous 
results without appealing to the somehow complicated U-duality 
transformations.  However, the entropy calculation in the charged case remains 
unsolved and suggests an open problem.  Through further investigation, it will 
be interesting to study the entropy in the GEMS approach to charged 
(1+1) dilatonic black holes and to investigate the relations between the GEMS 
and the U-duality schemes.

\acknowledgements{
STH would like to thank C. Nappi and I. Bars for helpful discussions and concerns.  
This work was supported by the Basic Research Program of the Korea 
Science and Engineering Foundation, Grant No. R01-2000-00015.}


\begin{references}
\bibitem{bhs} C. G. Callan, S. B. Giddings, J. A. Harvey, and
  A. Strominger, Phys. Rev. {\bf D45}, 1005 (1992); C-J. Ahn,
  W. T. Kim, Y-J. Park, K. Y. Kim, and Y. Kim, Mod. Phys. Lett. {\bf
  A7}, 2263 (1992); S. Kojima, N. Sakai, and Y. Tanii,
  Mod. Phys. Lett. {\bf A10}, 2391 (1995); W. T. Kim, J. Lee, and Y-J. Park, Phys. Lett. {\bf B347}, 217 (1995);  
T. Futamase, M. Hotta, Y. Itoh, Phys. Rev. {\bf D57}, 1129 (1998); 
I. Oda, Phys. Rev. {\bf D57}, 2415 (1998); 
W. Kummer, Annalen Phys. {\bf 8}, 801 (1999); S-W. Kim and H. Lee, 
J. Korean Phys. Soc. {\bf 35}, S666 (1999);
D. Youm, Phys. Rev. {\bf D61}, 044013 (2000);  
D. Grumiller, D. Hofmann, and W. Kummer, Annals Phys. {\bf 290}, 69 (2001);
S. Nojiri and S. D. Odintsov, Int. J. Mod. Phys. {\bf A16}, 1015 (2001).
\bibitem{witten91} 
G. Mandal, A. M. Sengupta, and S. R. Wadia, Mod. Phys. Lett. {\bf A6}, 1685 (1991); 
S. Elitzur, A. Forge, and E. Rabinovici, Nucl. Phys. {\bf B359}, 581 (1991);
E. Witten, Phys. Rev. {\bf D44}, 314 (1991).
\bibitem{nappi92} M. D. McGuigan, C. R. Nappi, and S. A. Yost, Nucl. Phys. {\bf B375}, 421 (1992).
\bibitem{nappi92mod} C. R. Nappi and A. Pasquinucci, Mod. Phys. Lett. {\bf A7}, 3337 (1992).
\bibitem{gibbons92} G. W. Gibbons and M. J. Perry, Int. J. Mod. Phys. {\bf D1}, 335 (1992). 
\bibitem{hyun98} S. Hyun, J. Korean Phys. Soc. {\bf 33}, S532 (1998), 
{\tt hep/th-9704005}.
\bibitem{hawk75} S. W. Hawking, Commun. Math. Phys. {\bf 42}, 199 (1975); 
J. D. Bekenstein, Phys. Rev. {\bf D7}, 2333 (1973); 
R. M. Wald, {\em Quantum Field Theory in Curved Spacetime and Black Hole 
Thermodynamics} (The University of Chicago Press, Chicago, 1994); 
J. D. Brown, C. Creighton, and R. B. Mann, Phys. Rev. {\bf D50}, 6394 (1994).
\bibitem{unr} W. G. Unruh, Phys. Rev. {\bf D14}, 870 (1976);
P. C. W. Davies, J. Phys. {\bf A8}, 609 (1975).
\bibitem{kasner} E. Kasner, Am. J. Math. {\bf 43}, 130 (1921); 
C. Fronsdal, Phys. Rev. {\bf 116}, 778 (1959); H. F. Goenner, 
{\em General Relativity and Gravitation} (Plenum, New York, 1980) Ed. A. Held;
 J. Rosen, Rev. Mod. Phys. {\bf 37}, 204 (1965); H. Narnhofer, I. 
Peter, and W. Thirring, Int. J. Mod. Phys. {\bf B10}, 1507 (1996).
\bibitem{deser97} S. Deser and O. Levin, Class. Quantum Grav. {\bf 14},
L163 (1997); Class. Quantum Grav. {\bf 15}, L85 (1998); Phys. Rev. 
{\bf D59}, 0640004 (1999).
\bibitem{des} M. Beciu and H. Culetu, Mod. Phys. Lett. {\bf A14}, 1 (1999);
P. F. Gonzalez-Diaz, Phys. Rev. {\bf D61}, 024019 (1999); 
L. Andrianopoli, M. Derix, G. W. Gibbons, C. Herdeiro, A. Santambrogio, and 
A. V. Proeyen, Class. Quant. Grav. {\bf 17}, 1875 (2000).
\bibitem{kps99} Y-W. Kim, Y-J. Park, and K. S. Soh, Phys. Rev. {\bf D62}, 
104020 (2000). 
\bibitem{kps00} S-T. Hong, Y-W. Kim, and Y-J. Park, Phys. Rev. {\bf D62}, 
024024 (2000); S-T. Hong, W. T. Kim, Y-W. Kim, and Y-J. Park, Phys. Rev.  
{\bf D62}, 064021 (2000).
\bibitem{btz1} M. Banados, C. Teitelboim, and J. Zanelli, Phys. Rev. 
Lett. {\bf 69}, 1849 (1992); M. Banados, M. Henneaux, C. Teitelboim, and
J. Zanelli, Phys. Rev. {\bf D48}, 1506 (1993).
\bibitem{cal} S. Carlip, Class. Quant. Grav. {\bf 12}, 2853 (1995).
\bibitem{kps} S. W. Kim, W. T. Kim, Y-J. Park, and H. Shin, Phys. Lett. 
{\bf B392}, 311 (1997).
\bibitem{mann93} D. Cangemi, M. Leblanc, and R. B. Mann, Phys. Rev. 
{\bf D48}, 3606 (1993).
\bibitem{jkps} J. Ho, W. T. Kim, and Y-J. Park,
J. Korean Phys. Soc. {\bf 33}, S541 (1998); H-C. Kim, Y. Kim, and P. Oh,
J. Korean Phys. Soc. {\bf 35}, S655 (1999); T. Lee, 
J. Korean Phys. Soc. {\bf 35}, S670 (1999).
\bibitem{sch} K. Schwarzschild, Sitzber. Deut. Akad. Wiss. Berlin, KI. 
Math.-Phys. Tech., pp. 189-196 (1916).
\bibitem{rn} H. Reissner, Ann. Physik {\bf 50}, 106 (1916);
G. Nordstrom, Proc. Kon. Ned. Akda. Wet {\bf 20}, 1238 (1918); 
C. S. Peca and J. P. S. Lemos, Phys. Rev. {\bf D59}, 124007 (1999); 
P. Mitra, Phys. Lett. {\bf B459}, 119 (1999); 
S. W. Hawking and H.S. Reall, Phys. Rev. {\bf D61}, 024014 (1999); 
B. Wang, E. Abdalla, and R-K. Su, Phys. Rev. {\bf D62}, 047501 (2000).
\bibitem{spivak75} M. Spivak, {\it Differential Geometry} (Publish or Perish, 
Berkeley, 1975) Vol 5, Chapter 11.
\bibitem{brane} M. Gogberashvili, gr-qc/0202061. 
\bibitem{hong01prd}  S-T. Hong, W. T. Kim, J. J. Oh, and Y-J. Park, Phys. 
Rev. {\bf D63}, 127502 (2001). 
\bibitem{horowitz962} G. T. Horowitz, J. M. Maldacena, and A. Strominger, 
Phys. Lett. {\bf B383}, 151 (1996).  
\bibitem{tsey96} A. A. Tseytlin, Mod. Phys. Lett. {\bf A11}, 689 (1996).
\bibitem{mal96} J. M. Maldacena, {\it Black holes in string theory}, Ph.D 
Thesis (Princeton University), hep-th/9607235.
\bibitem{berg95} E. Bergshoeff, C. Hull and T. Ortin, Nucl. Phys. 
{\bf B451}, 547 (1995). 
\bibitem{teo98} E. Teo, Phys. Lett. {\bf B430}, 57 (1998).
\bibitem{gib77} G. W. Gibbons and S. W. Hawking, Phys. Rev. {\bf D15}, 2738 
(1977); R. Laflamme, Phys. Lett. {\bf B196}, 449 (1987).
\bibitem{nature} S. W. Hawking, Nature {\bf 248}, 30 (1974). 
\end{references}
\end{document}